# Direct Observation of Acoustic Phonon Confinement in Free-Standing Semiconductor Nanowires

**Fariborz Kargar[1], Bishwajit Debnath[2], Kakko Joona-Pekko[3], Antti Säynätjoki[3,4], Harri Lipsanen[3], Denis Nika[1,5], Roger Lake[2,6] and Alexander A. Balandin[1,6,·]**

[1]Phonon Optimized Engineered Materials (POEM) Center and Department of Electrical and Computer Engineering, University of California – Riverside, Riverside, California 92521 USA

[2]Laboratory for Terascale and Terahertz Electronics (LATTE), Department of Electrical and Computer Engineering, University of California – Riverside, Riverside, California 92521 USA

[3]Department of Micro and Nanosciences, School of Electrical Engineering, Aalto University, FI-00076 Aalto, Finland

[4]Institute of Photonics, University of Eastern Finland, FI-80101 Joensuu, Finland

[5]Department of Physics and Engineering, Moldova State University, Chisinau, MD 2009, Republic of Moldova

[6]Spins and Heat in Nanoscale Electronic Systems (SHINES) Center, Bourns College of Engineering, University of California – Riverside, Riverside, California 92521 USA

**Similar to electron waves, the phonon states in semiconductors can undergo changes induced by external boundaries[1,2]. Modification of *acoustic* phonon spectrum in structures with *periodically* modulated elastic constant or mass density – referred to as phononic crystals – has been proven experimentally and utilized in practical applications[3–8]. A possibility of modifying acoustic phonon spectrum in individual nanostructures via spatial confinement would bring tremendous benefits for controlling phonon-electron interaction and thermal conduction at nanoscale[9–11]. However, despite strong scientific and practical importance, conclusive experimental evidence of *acoustic phonon confinement* in individual free-standing**

· Corresponding author (A.A.B.): balandin@ece.ucr.edu

nanostructures, e.g. nanowires, is still missing. The length scale, at which phonon dispersion undergoes changes and a possibility of the phonon group velocity reduction, are debated. Here, we utilize specially designed high-quality GaAs nanowires (NWs) with different diameters, *D*, and large inter-nanowire distances to *directly* demonstrate acoustic phonon confinement. The measurements conducted with Brillouin – Mandelstam spectroscopy reveal confined phonon polarization branches with frequencies from 4 GHz to 40 GHz in NWs with *D* as large as ~128 nm, i.e. at length scale, which exceeds the "grey" phonon mean-free path in GaAs by an almost an order of magnitude. The phonon dispersion modification and phonon energy scaling with *D* in individual nanowires are in excellent agreement with theory. The obtained results can lead to more efficient nanoscale control of acoustic phonons, with benefits for nanoelectronics, thermoelectric energy conversion, thermal management, and novel spintronic technologies.

[Main Text]

Engineering electron waves and energy dispersion with external boundaries in semiconductor heterostructures and quantum well superlattices became the foundation of modern electronics and optoelectronics, enabling high-electron mobility transistors, optical modulators and other devices[11]. Recent technological developments suggest that tuning phonon energy dispersion may become as important for the next generation of nanoelectronic circuits as engineering of the electron dispersion was for the first heterostructure-based devices. Acoustic phonons carry heat in semiconductors and electrical insulators. Improving phonon transport in nanometer scale devices is crucial for their thermal management and reliability. Phonons set performance limits for alternative technologies under development: from spintronics to quantum computing[12,13]. Continuous downscaling of a transistor size and emergence of heterostructures with elastically dissimilar materials have created conditions for tuning acoustic phonon spectrum in a desired way.

It has long been suggested theoretically that sound waves in layered media[14], or acoustic phonons in semiconductor superlattices and nanostructures, can undergo modification due to externally imposed periodic or stationary boundary conditions[15–17]. Modulation of the elastic constants, $C_{ij}$, or mass density, $\rho$, in such structures can lead to drastic changes in the acoustic phonon dispersion, $\omega(q)$, which include reduction in the phonon group velocity, $\upsilon_G = \partial\omega(q)/\partial q$ or emergence of the phonon energy band gaps (here $\omega$ and $q$ are the phonon energy and wave vector, respectively)[5,8,15–19]. There are many experimental reports of the phonon wave effects in *periodic* structures, including the early Raman spectroscopy studies in semiconductor superlattices[20]. The possibility of controlling the acoustic phonon spectrum in periodic structures has led to an explosive growth in the field of *phononic crystals*[4–8]. However, despite the strong motivations, and enormously large body of theoretical studies[10,11], conclusive experimental evidence of the *acoustic phonon confinement* in individual nanostructures (as opposed to periodic composites) is still missing.

Establishing the existence of modification of the acoustic phonon spectrum in free-standing nanostructures, and determining the length scale, $D$, at which the effects start to reveal themselves is extremely important. Controlling the phonon spectrum via spatial confinement would allow for fine-tuning of the phonon interactions with electrons, spins and other phonons. Commonly used assumption in the thermal community[21,22] is that the phonon spectrum modifications only appear when the structure size is on the order of the dominant thermal phonon wave length $\lambda_T \approx (\upsilon_s h)/(k_B T)$, which is ~1.5 nm for many solids at room temperature (RT) (here $k_B$ is the Boltzmann constant, $T$ is the absolute temperature, $h$ is the Planck constant, and $\upsilon_s$ is the sound velocity). This criterion is based on the notion that the phonon wave functions do not preserve coherence over larger distances owing to natural surface roughness in real nanostructures and crystal lattice inharmonicity. As a result, wave interference affects required for modification of the phonon dispersion do not take place even if solution of the elasticity equation predicts appearance of the confined phonon subbands. From the other side, a large body of literature[2,10,15,18] on phonons in semiconductors considers that the phonon spatial confinement effects start to take place when $D$ is on the order of the average phonon mean free path (MFP), defined in kinetic theory as the *grey* MFP: $\Lambda_G = 3K/(C_V \upsilon_s)$, which is ~20 nm for bulk GaAs at RT (here, $K$ and $C_V$ are the thermal conductivity and specific heat capacity, respectively). This is because even a small energy difference between the phonon subbands in the middle of the Brillouin zone (BZ) (phonon wave vector $q<10^6$ cm$^{-1}$) can result in changes in the electron – phonon scattering rates, and, correspondingly, electron momentum and energy relaxation rates, particularly at low temperature[2,12,23]. In the thermal context, if the long-wavelength phonons (with $\lambda>\lambda_T$) contribute to heat conduction, their spatial confinement should affect the thermal conductivity.

Recent studies established that the actual phonon MFP can be substantially larger than $\Lambda_G$[24–26]. It was found, contrary to conventional knowledge, that ~40% of thermal conductivity of crystalline Si near RT come from the phonons with MFP above 1 μm[26]. The phonons with the longest MFP may have the lowest frequency, $\omega$, extending down to GHz range. The phonon life-time scales as $\tau \propto \omega^{-s}$ (where $s \geq 1$ depends on the phonon scattering mechanism and dimensionality of the system; e.g. $s=4$ for point-defect scattering in bulk or nanowires with lateral dimensions larger than a few lattice constants). As a result, phonon MFP $\Lambda = \upsilon_G \tau$ will be growing with decreasing $\omega$. The discovery of the large contribution of the long-MFP phonons to thermal conductivity makes the task of determining the length scale of the onset of confinement effects even more important. The exceptionally large phonon MFP suggests that phonon wave may "feel" the boundaries of nanostructures over larger scales than previously expected.

Semiconductor NWs with smooth boundaries are the most suitable candidates for investigation of spatial confinement of acoustic phonons in the *free-standing* (or free-surface) nanostructures. For NWs, the characteristic length scale $D$ is the diameter. However, random distribution of diameters in typical NW arrays, small distances between NWs, surface roughness, and difficulty of direct detection of acoustic phonons so far precluded confirming the theoretically predicted phonon confinement effects in NWs[27,28]. A few published reports for NWs only infer possible confinement effects *indirectly* by measuring thermal conductivity as a function of temperature, and comparing the result with calculations, which use both bulk and confined phonon dispersion[29,30]. The latter approach is ambiguous owing to the difficulty of separating the phonon confinement effects from the phonon – rough interface scattering. The Raman spectroscopy studies of the phonon zone folding[31,32] or phonon energies in the BZ center of the layered van der Waals materials[33,34] are important developments in the field but they cannot be readily extended to conventional semiconductors and nanowire geometries. The Raman studies do not provide phonon dispersion, which is required for proving the phonon group velocity reduction in real nanostructures.

In this Letter, we report measurements of acoustic phonon spectrum of NWs in a unique set of NW arrays with different $D$, which allowed us to *conclusively* prove the existence of spatial phonon confinement effects in individual nanostructures. We have also discovered that the effects become pronounced at substantially larger length scale than previously believed. The measurements are conducted in GHz range of phonon frequencies, which dominates heat conduction at low temperature[35]. The fact that we established the existence of the confined phonon branches at RT indicates that the effects are even stronger at low temperature. While THz phonons make the main contribution to thermal transport at RT, GHz phonons participate in the phonon – phonon scattering processes[1,36]. The role of the low-frequency phonons is greatly increased in alloyed materials because the higher frequency phonons scatter stringer[37]. The most recent experimental confirmation of the dominant role of phonons with the extra-long MFP, even at RT, further explains the relevance of the examined phonon frequency range to heat conduction. The BZ-center phonons ($q < 10^6$ cm$^{-1}$) are essential for controlling electron relaxation[2,15–17,23].

Several sets of specially designed GaAs NWs with excellent surface quality and hexagonal cross section have been fabricated on top of GaAs (111) substrate by the selective-area epitaxy (SAE) utilizing the metal-organic vapor phase epitaxy (MOVPE) system. The top facets of NWs have the crystallographic plane same as the substrate while the side-facets are {110} planes. The NWs were grown in a perfect vertical arrangement in hexagonal arrays with the diameter ranging from 103 nm up to 135 nm. The important attributes of the samples were (i) diameter uniformity within each batch (the relative standard deviation was ~3% for most tested NWs), (ii) large distance, $H$, between the NWs (we focused on the range from 700 nm to 3 μm; the smallest $H$ was 250 nm while the largest was 10 μm), and (iii) large length, $L$, of NWs (at least 10 times of the NWs diameter). A scanning electron microscopy (SEM) image of a representative sample is shown in Figure 1 (a). Details of the sample fabrication and characterization are provided in the *Methods* and *Supplementary Information*.

We have used Brillouin-Mandelstam light scattering (BMS) spectroscopy as a tool to measure dispersion of acoustic phonons with energies in the range from ~2 GHz up to 200 GHz near the Brillouin zone (BZ) center. Various modifications of Brillouin spectroscopy[38,39] are gaining popularity for investigating acoustic phonon and magnon energies in phononic crystals and other materials[40–50]. Changing the angle of light incidence, $\alpha$, with respect to the substrate allowed us to vary the probing phonon wave-vector, $q$, and determine the dispersion near the BZ center. A schematic of the experiment is shown in Figure 1 (b). The large inter-NW distance $H$, which was much larger than the laser wavelength, $\lambda$, ensured that NWs scatter light individually, e.g. neither light interference nor elastic coupling via the substrate affect the results. We used the samples with $H$ up to ~3 μm to verify that the measured spectral characteristics are independent of $H$. Large $H$ and $L>>D$ also ensure that each NW can be modeled theoretically as an infinite NW with free-surface boundaries. These unique characteristics of NW samples made possible investigation of the phonon spectrum features originating in individual NWs.

There are three mechanisms, which contribute to light scattering in our samples. They are scattering (i) from the bulk, i.e. the volume of the substrate via the elasto-optic mechanism[38,39], (ii) from the surface of the substrate via the *surface ripple* mechanism[38,39], and (iii) from the side-facets of NWs via the surface ripple mechanism as well. In the volumetric mechanism (i) the phonon wave vector, $q$, is fixed at $q_B = 4\pi n/\lambda$ where $n$ is the refractive index of GaAs. In the ripple scattering mechanisms (ii) and (iii), the phonon wave vector is given by $q_{S-S} = (4\pi/\lambda)\sin(\alpha)$ for the substrate and $q_{S-NW} = (4\pi/\lambda)\cos(\alpha)$ for NWs (see notation for angle $\alpha$ in Figure 1 (b)). The difference in the angle dependence is due to the perpendicular alignment of NWs. In probing the ripple scattering mechanism, changing $\alpha$ allows one to vary $q$ within a certain range. The volumetric elasto-optic scattering is absent in our NWs because $D < \lambda$. Although the light scattering is from the NW surface ripples it inherently depends on the confined phonon modes inside NW, which create the ripples. In our experiments, changing $\alpha$ from 15° to 40° corresponded to 6.1 μm⁻¹ ≤ $q_{S-S}$ ≤ 15.2 μm⁻¹ and 18.1 μm⁻¹ ≤ $q_{S-NW}$ ≤ 22.8 μm⁻¹ in the ripple scattering from the substrate and NWs, respectively. Taking into account that $n$=4.13 for GaAs at $\lambda$=532 nm the $q$ value for the substrate elasto-optic scattering was fixed at $q_B$=97.6 μm⁻¹.

Figure 1 (c) shows BMS spectra of NWs with the diameter $D$=122 nm (inter-NW distance $H$=800 nm) extracted from SEM data, and the substrate without NWs in the frequency range from 20 GHz to 125 GHz. The phonon peaks at 45.6 GHz and 85.8 GHz in the spectrum from the substrate correspond to the transverse acoustic (TA) and longitudinal acoustic (LA) polarization branches. These peaks originate from the *true acoustic* bulk phonons that have zero frequency at the BZ center, i.e. $\omega(q=0)$=0. The bulk phonon peaks are present in the spectra of NW samples as well because part of the signal is coming from the substrate. The most interesting feature of NW spectrum is the appearance of additional peaks attributed to the confined acoustic (CA) phonons. These phonons are quasi-optical in nature, in a sense, that their energy is non-zero at the BZ center, i.e. $\omega(q=0)$≠0. All confined phonons in NWs are hybrid in nature comprising of vibrations with different polarization. Although confined phonons are higher in energy than the true acoustic phonons, in the experimental spectrum, they appear at smaller frequencies because the probing phonon wave vectors are different: $q_{S-NW}$<$q_B$. Figure 1 (d) presents the evolution of the BMS spectrum for the same NW sample for different values of $q$ that are varied by changing the angle $\alpha$. We were able to resolve six confined phonon branches denoted as CA$_i$ ($i$ is the confined phonon index). The peak below 15 GHz is a mixture of the true LA phonons in the substrate and LA-like phonons in NWs.

[Figure 1]

To understand and confirm the confined nature of the observed phonon peaks above 15 GHz, we solved the elasticity equation for NWs using the finite-element method. The simulations were carried out for NWs with hexagonal cross-sections using GaAs elastic constants for a specific crystallographic direction ([111]) and NW diameter determined from SEM. Figure 2 (a) shows the simulated phonon dispersion (solid curves) in a hexagonal NW along the [111] direction for $D$=122 nm. One can see excellent agreement between the experimental data and modeling results for both true acoustic and confined phonon branches. Experimental uncertainties, on the order of ~1 GHz, is within the standard deviation of the NW diameters (i.e., 2.9% for $D$=122 nm). Other sources of uncertainty are inevitable variations in the elastic behavior of NWs resulting from the stacking faults and possible small inclusions of the wurtzite (WZ) phase to the dominant zinc-blend (ZB) polytype in NWs with various diameters. This may result in a small change in the LA-like sound velocity $v_s = (C_{11}/\rho)^{1/2}$, defined by the slope of this branch. Additional analysis of the model sensitivity to parameter variations is provided in *Supplementary Materials*.

Our proof of the confined nature of the phonons is based on excellent agreement of the dispersions obtained from BMS experiments and calculated for the exact NW shape and material parameters. In order to confirm that the identified phonon modes are Brillouin active, we calculated the average surface displacement of NW's side-facet perpendicular to $q_{S\text{-}NW}$. In the surface scattering mechanism, the phonon modes that produce displacement perpendicular to $q_{S\text{-}NW}$ in the plane of scattering are those likely to contribute to the BMS spectrum. The scattering cross-section is given by[51]

$$d^2\sigma/d\Omega\,d\omega_s = \left(\zeta\;\omega_I^4/16\pi^2\,c^4\right)F^2\left\langle\left|u^z(0)\right|^2\right\rangle_{q_x,\omega} \tag{1}$$

Here $\zeta$ is a coefficient proportional to the illuminated area, $\omega_I$ is the frequency of the incident light, $c$ is the speed of light in the vacuum, $F$ is a function of the incident and scattered light angles as well as $n$ of the scattering medium. The last term $<\ldots>$ is the surface-displacement power spectrum with $q_x$ projection to the surface. The calculated displacements, which enter Eq. (1), are shown in Figure 2 (b) by the color of the simulated phonon branches. The results are in agreement with the experiment in a sense that the measured data points are all within the Brillouin active segments of the polarization branches. The fact that these branches are active is in line with the observation that the considered phonon modes are hybrid in nature.

The phonon modes are visualized in Figures 2 (c) and (d) as the normalized displacement field distributions in NWs for two wave-vectors $q_{S-NW}$=0.3 μm$^{-1}$ and $q_{S-NW}$=18.0 μm$^{-1}$ – close and away from the BZ center, correspondingly. The modes become more hybrid in nature as $q$ increases. The active modes show strong surface ripple on the side facets of NWs. The symmetry of the confined modes is complicated. The point group of ZB is $T_d$ with 24 invariant operations whereas WZ is characterized by C$_{6v}$[52]. Confined NW geometry reduces the symmetry to C$_{3v}$. The two lowest branches are flexural acoustic (FA) modes of E$_1$ symmetry with the quadratic dispersion $\omega \propto k^2$. The degeneracy of this branch is broken for NWs along the [111] direction. The next two acoustic modes, torsional acoustic of A$_2$ symmetry and LA-like of A$_1$ symmetry, respectively, have linear dispersion $\omega \propto k$, near the BZ center. The confined branches mainly belong to the higher symmetry groups *e.g.* E$_2$, B$_1$, B$_2$. See animation of the modes in *Supplemental Video*.

[Figure 2]

The final element of evidence of spatial confinement of acoustic phonons in free-standing NWs comes from the analysis of the diameter dependence of the CA phonon frequencies (energies). Figure 3 (a) shows the measured phonon spectra for different NW diameters, $D$ ($H$=800 nm for all samples). One can see that with increasing $D$, the frequencies of the confined phonons decrease. The trend and the magnitude of the frequency change are consistent with the theory (see calculated dispersions for various $D$ in the *Supplementary Information*). In Figure 3 (b) we present spectra of NWs with the constant $D$=122 nm and varying $H$. The spectral position of CA peaks does not depend on the inter-NW distance. The absence of $H$ dependence indicates that the measured spectral features are characteristics of *individual* NWs. There is no elastic coupling among NWs. The spatial phonon confinement in the free-surface NWs is distinctively different from the phonon spectrum changes owing to the periodic boundary conditions in the phonon band-gap materials.

[Figure 3]

A remaining intriguing question in confirming the confined nature of CA peaks is why their full-width-at-half-maximum (FWHM) appear smaller than that of the substrate LA phonon peak (see Figure 1 (c)). Theory suggests that the phonons with smaller group velocity, $v_G$, should be more strongly scattered[18]. The phonon life-time limited by the point-defect scattering, $\tau_D = (4\pi v_G^3)/(V_o \omega^4 \Gamma)$, rapidly decreases with decreasing $v_G$ (here $V_o$ is the volume per atom and $\Gamma$ is the defect scattering factor). The Umklapp limited phonon life-time also decreases with decreasing $v_G$. The answer to this question is that FWHM of the elasto-optic and ripple mechanism peaks cannot be compared directly. The FWHM of the LA peak from the substrate ($\Delta\omega = 13.6$ GHz) is defined by the light absorption. For an opaque crystal with the refractive index $n = n_1 + i n_2$, theoretical broadening for the elasto-optic scattering[53] is $\Delta\omega/\omega = 2 n_2/n_1$. Using for GaAs, $n_1$ =4.13 and $n_2$=0.34[54], we obtain $\Delta\omega/\omega = 0.164$, which closely matches with the measured $\Delta\omega/\omega$ =0.158. For the CA phonons, observed in our experiments via the surface ripple scattering, the peak broadening is defined by the aperture effects.[38,39]

The discovery that the acoustic phonon spectrum becomes strongly modified near BZ center at the length-scale much larger than the *grey* phonon MFP or the thermal phonon wavelength has important fundamental science and practical implications. It means that the acoustic phonon confinement can affect the electron – phonon scattering rates[55] in the structures comparable in size to the state-of-the-art electronic devices. The frequency-dependent average phonon group velocity[56] $\langle v(\omega) \rangle = g(\omega)/\Sigma_s 1/v_s(\omega)$, which affects the heat conduction, also changes (here $g(\omega)$ is the number of phonon branches in the spectra with the frequency $\omega$, and $v_s(\omega)$ is the group velocity of phonon mode $(s, f)$). In NWs with $D$~100 nm, $\langle v(\omega) \rangle$ drops to less than half of the LA-like mode sound velocity in the frequency range around 20 GHz – 60 GHz due to emergence of the confined phonon branches.

## Acknowledgements

The work at UC Riverside was supported as part of the Spins and Heat in Nanoscale Electronic Systems (SHINES), an Energy Frontier Research Center funded by the U.S. Department of Energy, Office of Science, Basic Energy Sciences (BES) under Award # SC0012670. The sample preparation at Aalto University was supported by Aalto University's Energy Efficiency project Moppi and by Tekes NP-NANO project. K.J.P. acknowledges the support of Aalto University's ELEC Doctoral School. The authors thank E. Hernandez and Z. B. Beiranvand for help with images. A.A.B. acknowledges useful discussions on BMS instrumentation with Prof. C. M. Sotomayor Torres and Prof. Xiaoqin Li.

## Author contributions

A.A.B. conceived the idea, coordinated the project and led the experimental and theoretical data analysis; F.K. conducted BMS measurements, material characterization and contributed to data analysis; B.D. performed numerical simulations; R.L. supervised numerical simulations and contributed to data analysis; K.J.P. and A.S. fabricated the nanowire samples and carried out material characterization; H.L. supervised nanowire synthesis and contributed to material characterization; D.N. contributed to theoretical data analysis. A.A.B. wrote the manuscript. All other authors contributed to preparation of a final manuscript.

## Additional information

Supplementary information accompanies this paper at www.nature.com. Reprints and permission information is available online at http://npg.nature.com/reprintsandpermissions/. Correspondence and requests for materials should be addressed to [A.A.B.]

**METHODS**

**Sample preparation:** GaAs NWs were fabricated with SAE using MOVPE system (Thomas Swan)[57]. The growth was performed on p-type GaAs (111) B substrates in atmospheric pressure. Prior to NW growth the growth templates were fabricated as follows. First, a 40-nm thick $SiO_x$ layer was deposited using the plasma enhanced chemical vapor deposition (PECVD) (Oxford Systems). Next, electron beam lithography (EBL) (Vistec) was performed to pattern the growth templates and reactive-ion etching (RIE) was used to transfer the patterns to the deposited $SiO_x$ layer. Poly(methyl methacrylate) (PMMA) (MicroChem) was used as the EBL resist. The patterns contained triangular lattice arrays of circles with different diameters and pitches. Each array of circles was 100 μm by 100 μm in size and the diameter of the circles was varied from 40 nm up to 125 nm with a 5 nm step. The pitch was varied from 250 nm up to 10 μm with minimum step size of 50 nm. Prior to transferring the samples into the MOVPE reactor, resist stripping, degreasing and cleaning were performed in acetone, isopropanol and de-ionized water. The NW growth was performed using trimethylgallium (TMGa) and tertiary-butyl arsine (TBAs) as the precursors with $H_2$ carrier gas and total gas flow of 5 slm. The molar flows were 0.811 µmol/min and 226.1 µmol/min for TMGa and TBAs, respectively. The samples were thermally cleaned at 760°C under TBAs flow for 5 min just before initiating NW growth by turning TMGa flow on. The growth temperature was the same as for the thermal cleaning. Three samples were fabricated with different NW growth time. The growth times were 6 min, 11 min 30 s and 15 min. This was done to acquire approximately equal height for smaller and larger diameter NWs. When the growth was finished, the TMGa flow was cut off and the samples were cooled to 150°C under TBAs protection.

**Brillouin-Mandelstam spectroscopy:** The experiments were carried out in "*p*-unpolarized" backscattering geometry using a solid-state diode pumped laser operating at $\lambda$=532 nm. The laser light was focused on the samples through a lens with *NA*=1.4. The scattered light was collected with the same lens and directed to the high-resolution six-pass tandem Fabry-Perot interferometer (JRS Instruments). A specially designed stage allowed to rotate the samples up to 60° relative to the direction of the incident laser light with an accuracy of 0.02°.

**Finite-element method simulations:** The phonon dispersion and displacement patterns have been calculated in the elastic continuum approximation using FEM implemented in COMSOL Multiphysics package. The GaAs sample is assumed to have zinc-blend (ZB) crystal structure. The out-of-plane direction of the nanowire is along x, which is also the growth direction (*i.e.* [111]). From the second-order continuum elastic theory, the equation of motion for the elastic vibration is

$$\rho \left( \partial^2 u(r) / \partial t^2 \right) = \partial S(r) / \partial x_i \tag{2}$$

Here u(**r**) is the three-component displacement vector at coordinate **r**; $\rho$ is the mass density; $S(\mathbf{r})$ is stress tensor that can be constructed from displacement by $S_{ij} = C_{ijkl}\ \varepsilon_{kl}$ with the six-component elastic strain tensor, $\varepsilon(\mathbf{r})=\frac{1}{2}[(\nabla\mathbf{u})^T+(\nabla\mathbf{u})]$[58]. The fourth-ranked elastic stiffness tensor, $C_{ijkl}$, is also expressed in the non-tensor notation as $C_{ij}$ with indices i, j, k, l running over the spatial coordinates (x, y, z). The elastic constants, $C_{ij}$, used in simulation correspond to GaAs in [001] direction: $C_{11}$=118.8 GPa, $C_{12}$=53.8 GPa, $C_{44}$=59.4 GPa[52]. According to *ab initio* simulations[52] and experiment[52], the bulk stiffness tensor is equally applicable to nanowire geometry as long as the nanowire diameter is not smaller than 20 nm. In order to take into account the growth direction of the nanowire we applied the tensor rotation operation to transform the stiffness matrix in [111]-oriented coordinate system as

$$C_{ijkl}^{[111]} = \Sigma_{\alpha\beta\gamma\delta} U_{i\alpha} U_{j\beta} U_{k\gamma} U_{l\delta} C_{\alpha\beta\gamma\delta}^{[001]} \tag{3}$$

Here $U$ is the rotational matrix for the Euler angles. The transformed elastic matrix and material parameters are summarized in Table I. The simulation geometry is discretized using finite element scheme to obtain the solution of the elasticity equation in the frequency domain as- $\rho\,\omega^2 u = \nabla.S$, where $\omega$ is the Eigen frequency. The free surface boundary conditions are applied at all outer facets in the radial direction by setting $\varepsilon_{ij}n_j$=0, where $n_j$ is the outward normal unit-vector.

**Table I: Material Parameters**

| $C_{11}$ (GPa) | $C_{12}$ (GPa) | $C_{13}$ (GPa) | $C_{15} = -C_{25} = -C_{46}$ (GPa) | $C_{33}$ (GPa) | $C_{44}$ (GPa) | $C_{66}$ (GPa) | $\rho$ (kg/m$^3$) | $\varepsilon_r$ |
|---|---|---|---|---|---|---|---|---|
| 145.7 | 44.8 | 35.8 | 12.7 | 154.6 | 41.5 | 50.4 | 5317 | 13.18 |

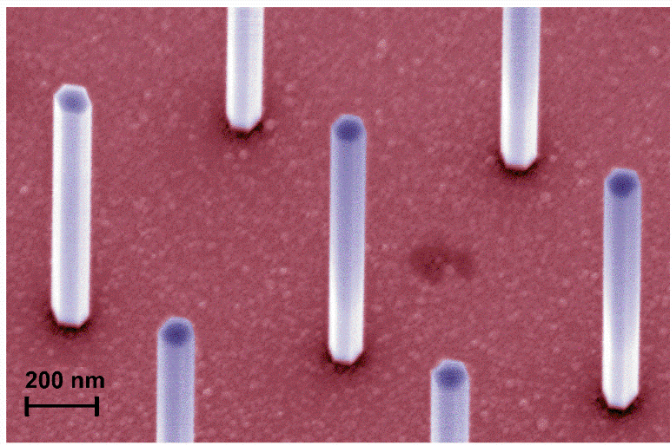
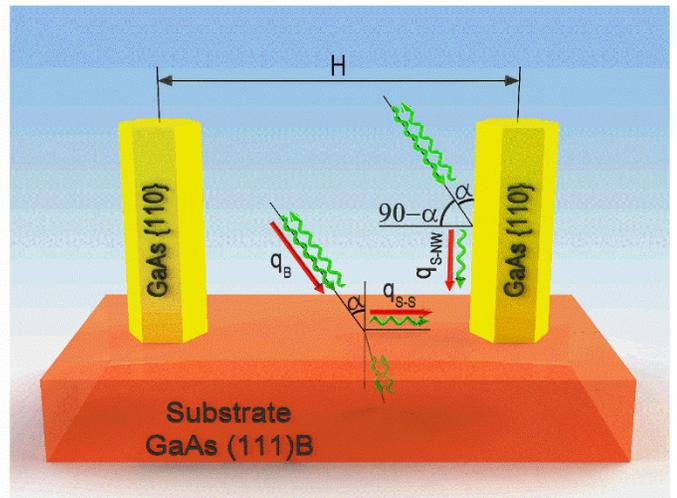

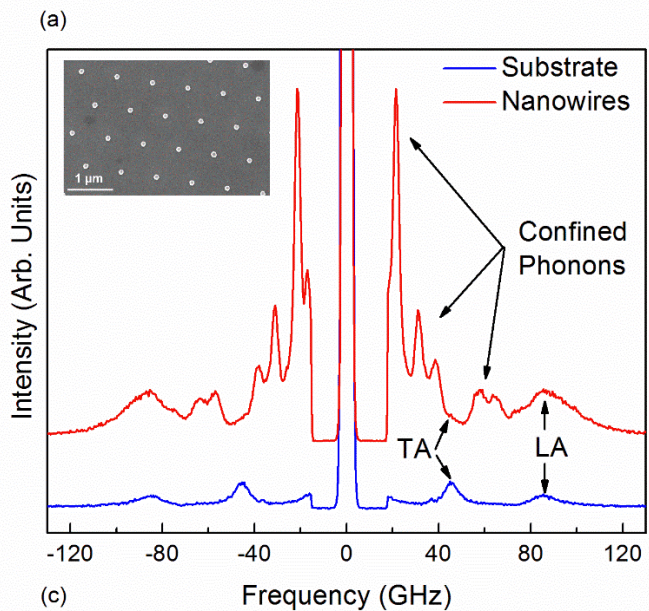
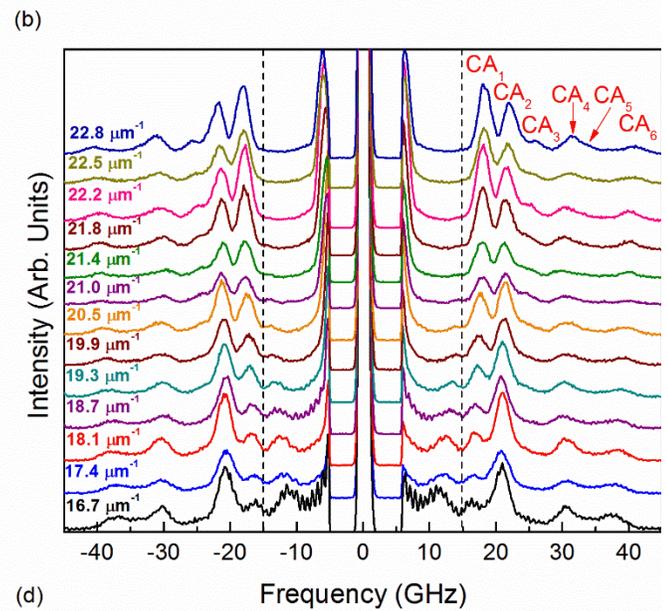

**Figure 1: Free-standing GaAs nanowires and their acoustic phonon spectrum.** (a) SEM image of NWs showing their diameter uniformity, orientation and large inter-nanowire distances (reaching up to $H$=3.0 μm in some samples). The pseudo-colors are used for clarity. (b) The schematic of the BMS experiment with notation for the substrate angle of incident $\alpha$, which translates to $\pi/2\text{-}\alpha$ for NWs. Two mechanisms of light scattering – elasto-optic inside the substrate and ripple scattering from the side-facets of NWs and substrate surface – are illustrated with green arrows. (c) Measured phonon spectrum for NWs with the diameter $D$=122 nm (red curve) and a substrate without NWs (blue curve). The inset shows a top-view SEM image of a representative NW sample. The regular LA and TA phonon peaks are present in both spectra. Additional peaks in the NW spectrum correspond to the confined acoustic phonons in individual NWs visible via the ripple scattering mechanism. These peaks appear at lower frequency than bulk phonons owing to the difference in the probing phonon wave vector $q$. (d) Evolution of the spectrum with the changing probing phonon wave vector defined by $\alpha$. The confined phonon branches are denoted as CA₁, CA₂, etc. The spectral position of the peaks was accurately determined by the Lorentzian fitting.

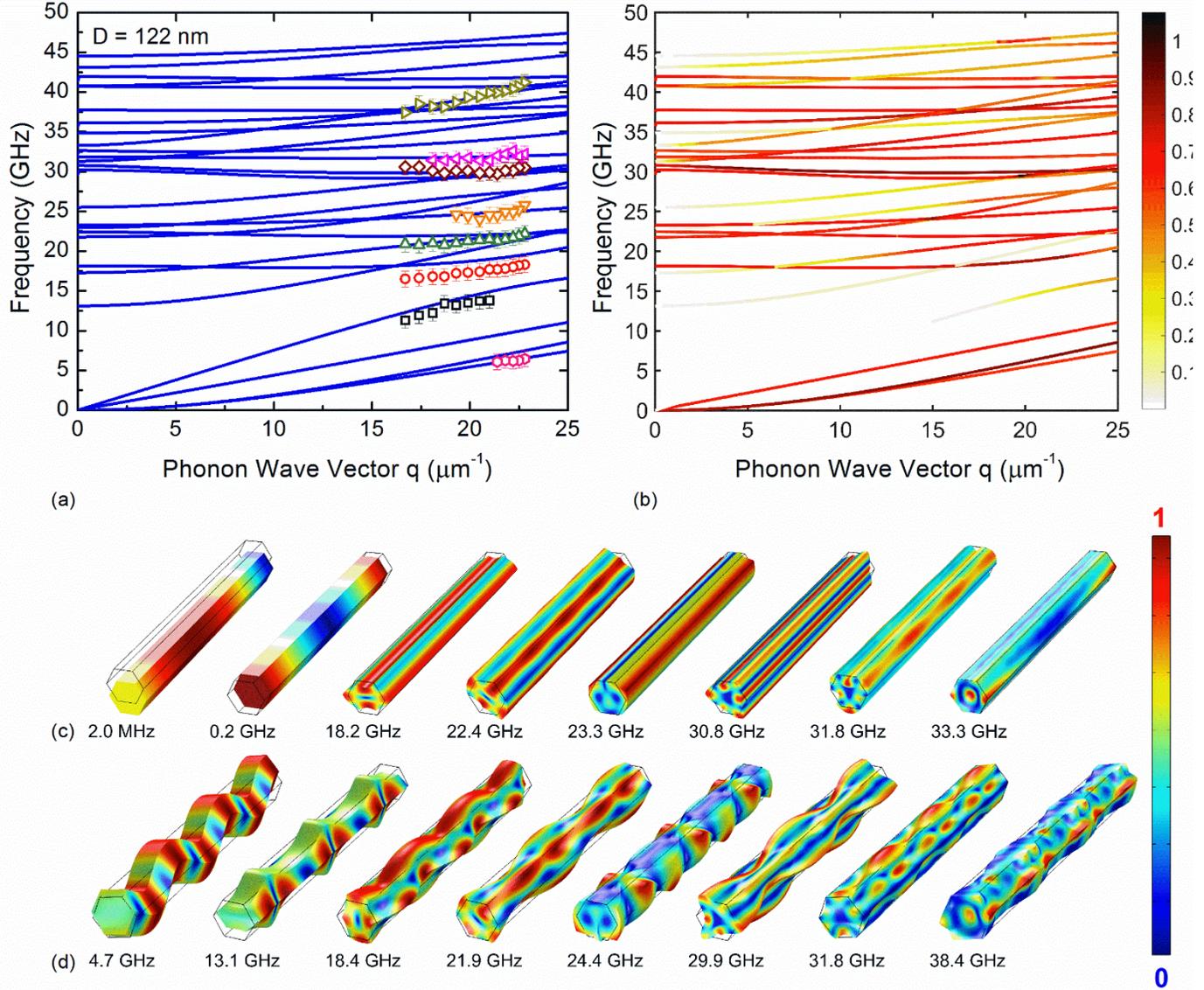

**Figure 2: Confined acoustic phonon dispersion in semiconductor nanowires.** (a) Measured and calculated phonon dispersion for a GaAs nanowire along [111] direction. The experimental data points, indicated with symbols, were obtained for NWs with the diameter $D$=122 nm determined from SEM inspection. The experimental data points are indicated with symbols and error bars. (b) Calculated dispersion with the color indicating the average surface displacement of the NW side-facet perpendicular to the phonon $q_{S-NW}$. Darker color corresponds to higher phonon mode activity in light scattering. (c) The normalized displacement field of the Brillouin-active phonon modes calculated for a 1-μm long NW for (c) $q_{S-NW} = 0.3$ μm⁻¹ and (d) $q_{S-NW} = 18.0$ μm⁻¹. The red color corresponds to stronger displacement.

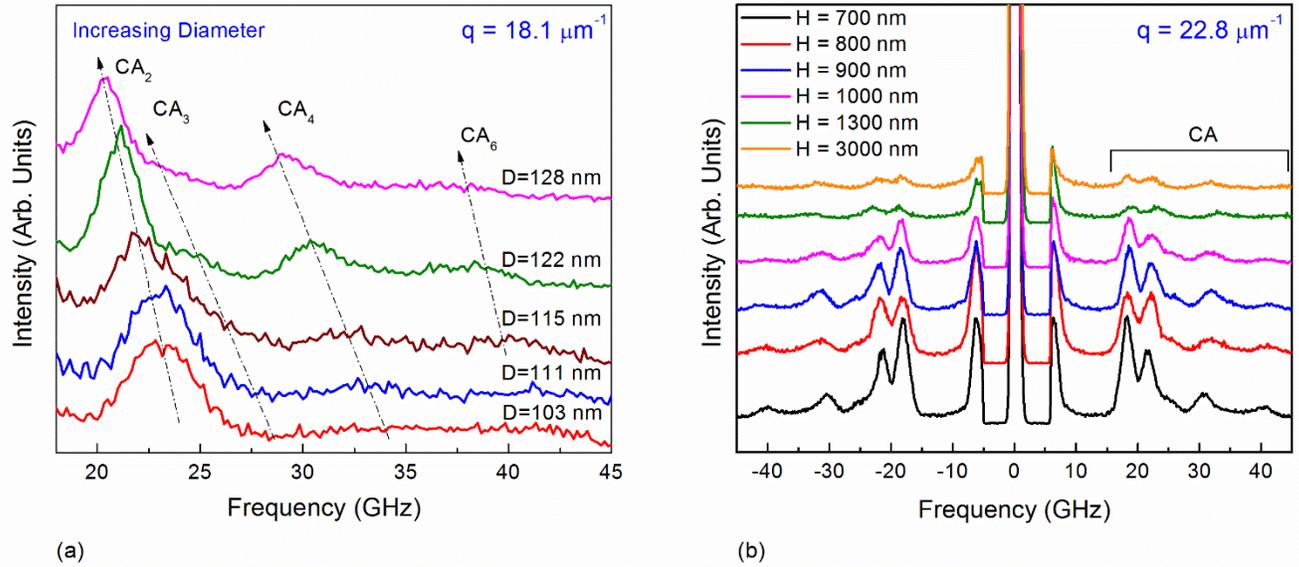

**Figure 3: Confined phonon energies in nanowires: effect of the diameter and inter-nanowire distance.**
(a) Brillouin-Mandelstam spectrum for NWs with different diameter at a constant probing phonon wave vector $q_{S-NW}$=18.1 $\mu m^{-1}$. The decrease in the frequency of the confined phonons with increasing NW diameter is clearly visible. The confined phonon branches show strong diameter dependence even for relatively large $D$ values in the range from ~103 nm to ~128 nm. The diameter dependence proves conclusively the presence of the spatial confinement effects at the length scale above the "grey" phonon mean-free path. (b) Measured spectrum for a set of NWs with the constant diameter $D$ = 122 nm and varying inter-NW distance $H$. The data are presented for the same fixed accumulation time of 30 minutes. The spectral position of the CA peaks does not depend on $H$. The intensity decreases with increasing $H$ owing to smaller number of illuminated NWs. The absence of the inter-NW distance dependence proves that the observed spectral features are characteristics of individual NWs.